\documentclass[12pt,leqno]{article}
\usepackage{latexsym}
\newtheorem{theorem}{Theorem}

\newtheorem{lemma}{Lemma}

\newtheorem{definition}{Definition}

\newenvironment{notation}{\noindent\bf Notation:\em\penalty100}{}

\newcommand{\blackslug}{\mbox{\hskip 1pt \vrule width 4pt height 8pt 
depth 1.5pt \hskip 1pt}}
\newcommand{\qed}{\quad\blackslug\lower 8.5pt\null\par\noindent}
\newenvironment{proof}{\par\noindent{\bf Proof:}}{\qed \par}

\newcommand{\cH}{\mbox{${\cal H}$}}

\title{A presentation of Quantum Logic based on an {\em and then} connective
\thanks{This work was partially supported 
by the Jean and Helene Alfassa fund for 
research in Artificial Intelligence and by the Israel Science Foundation grant 
183/03 on ``Quantum and other cumulative logics''}
}

\author{Daniel Lehmann\\Selim and Rachel Benin School of \\Computer Science and
Engineering, 
\\Hebrew University, \\Jerusalem 91904, Israel
}
\date{January 2007}

\begin{document}
\maketitle
\begin{abstract}
When a physicist performs a quantic measurement, new information about
the system at hand is gathered. This paper studies the logical properties
of how this new information is combined with previous information. It
presents Quantum Logic as a propositional logic under two connectives: 
negation and the {\em and then} operation that combines old and new 
information. The {\em and then} connective is neither commutative nor
associative. Many properties of this logic are exhibited, and some small
elegant subset is shown to imply all the properties considered.
No independence or completeness result is claimed. Classical physical
systems are exactly characterized by the commutativity, the associativity, 
or the monotonicity of the {\em and then} connective.
Entailment is defined in this logic and can be proved to be a partial order.
In orthomodular lattices, the operation proposed by Finch 
in~\cite{Finch_lattice:69} satisfies all the properties studied in this paper.
All properties satisfied by Finch's operation in modular lattices are valid
in Quantum Logic.
It is not known whether all properties of Quantum Logic are satisfied
by Finch's operation in modular lattices. Non-commutative, non-associative
algebraic structures generalizing Boolean algebras are defined, ideals
are characterized and a homomorphism theorem is proved. 
\end{abstract}

\section{Introduction} \label{sec:intro}
\subsection{Background} \label{sec:background}
Since its foundation in~\cite{BirkvonNeu:36}, an impressive
amount of different systems have been proposed for Quantum Logic.
This paper proposes a minimalistic syntax: 
one unary, $\neg$, and one binary, $*$, connectives.
The binary connective is not the commutative and associative conjunction
proposed by Birkhoff and von Neumann but the non-commutative, non-associative
conjunction proposed by Finch in~\cite{Finch_lattice:69} that is interpreted
in this paper as an {\em and then} connective acting on 
experimental propositions.
The minimalistic syntax provides algebraic properties that have an
immediate meaning for the logic of measurements in Quantum (and classical) 
Physics. Central properties of interest are properties of the binary
connective, $*$, alone, that do not mention $\neg$.
The algebraic structures, NCNAB-algebras, 
that correspond to this Quantum Logic are non-commutative, 
non-associative algebras that generalize Boolean algebras. 
The algebraic properties of the conjunction define 
an orthomodular partial order on the elements.
The commutative NCNAB-algebras are exactly the Boolean algebras, 
fitting the accepted wisdom that Classical Physics is the special case
of Quantum Physics one obtains when all observables commute.  
 
This should be contradistincted with traditional presentations of Quantum Logic
which:
\begin{itemize}
\item use a syntax including one unary connective and at least two binary
connectives: conjunction, disjunction and often one or more implications,
\item interpret conjunction as the (commutative) intersection of closed
linear subspaces of Hilbert space, which is semantically problematic since
the projection on the intersection $A \cap B$ of two closed subspaces cannot be
defined using the two projections on $A$ and $B$,
\item leads to a presentation in which the central properties considered
such as distributivity, modularity or orthomodularity involve more than one
connective, and have no obvious meaning for proof-theory.
\end{itemize}

Previous work on the non-commutative conjunction proposed by 
Finch~\cite{Finch_lattice:69}, such as~\cite{Roman_Rumbos:91} have
always considered this connective as defined 
in terms of more basic connectives.
This paper is closely connected to~\cite{LEG:Malg}.
The main difference is that, there,
the basic operation was composition of projections
and, here, the basic operation is the projection of one closed subspace 
on a closed subspace.

This paper leaves many questions unsolved.

\subsection{Plan of this paper} \label{sec:plan}
In Section~\ref{sec:what?} the formal framework of Quantum Mechanics 
is presented and the representation of knowledge about a quantic system
in this framework is discussed.
Section~\ref{sec:syntax} presents the syntax of the language 
that will be used to talk about quantic systems.
Section~\ref{sec:semantics} presents a semantic account of this language and
defines Hilbert Space Quantic Logic. 
Section~\ref{sec:ncb-alg} defines the corresponding first-order
structures, called NCNAB-algebras. 
They generalize Boolean algebras. It provides an in-depth
study of NCNAB-algebras.
Section~\ref{sec:orthomod} shows that orthomodular lattices, under
Finch's~\cite{Finch_lattice:69} interpretation of the {\em and then}
connective satisfy a list of central properties of NCNAB-algebras.
All properties of even modular lattices, under this interpretation of the
{\em and then} connective, hold in NCNAB-algebras.
Section~\ref{sec:homomorphism} studies ideals in NCNAB-algebras and proves
a homomorphism theorem.
Section~\ref{sec:future} is a summary and conclusion. 

\section{What is a quantic proposition?} \label{sec:what?}
When, in~\cite{BirkvonNeu:36}, Birkhoff and von Neumann 
introduced Quantum Logic,
they argued that an {\em experimental proposition} must be mathematically
represented by a closed (linear) subspace of a Hilbert space. 
Let us develop this point.

The formalism generally accepted for Quantum Mechanics, 
brought to its final form by von Neumann in~\cite{vonNeumann:Quanten}, 
considers the set of possible
states of a system to be the rays (i.e., one-dimensional subspaces) of 
a Hilbert space, say \cH.
A fundamental principle of Quantum Mechanics claims that if, 
from all one knows, the system could be in any one of two different states,
then it could be in any one of the many different superpositions of those
two states. Therefore propositions must be represented by linear 
subspaces of \cH.
Birkhoff and von Neumann argue that such subspaces must be
closed. 

Their argument is essentially the following.
The basic pieces of information one can gather about a system are of the type:
{\em the system is in the eigensubspace of some self-adjoint operator for
some eigenvalue $\lambda$}. The eigensubspaces of any bounded linear operator
are closed, and self-adjoint operators are bounded.
Then they explain that the information one can gather about any system 
is built out of those basic pieces by intersection 
(for information given by different commuting operators)
and linear sum (for different possible eigenvalues). 
They argue that, even for
infinite such sums, the result has to be understood as the closure 
of the linear span of the closed subspaces considered.

If a proposition is represented by a closed subspace $A$, one may, at least
in principle, test the system for this proposition.
The measurement, represented by
the projection on $A$ will, if the system is in a state that
satisfies the proposition (i.e., in $A$), give the corresponding eigenvalue
with probability one and, if the system does not satisfy the proposition, the
measurement will give, with some strictly positive probability, some other
eigenvalue.  

Consider now a totally unknown system on which one performs a sequence of 
two measurements. Before the first measurement, our knowledge is represented
by the whole (closed) space \cH.
After the first measurement, our knowledge is represented by the closed
subspace $A$ that is the eigensubspace corresponding to the result obtained.
After the second measurement, one knows, not only that the system is in
the closed subspace $B$ corresponding to the result obtained in the second
measurement, but also that it is in the projection on $B$ of some ray of $A$.
We must therefore consider that, if $A$ and $B$
are meaningful closed subspaces, then the projection of $A$ on $B$,
i.e., the direct image of $A$ under the transformation $\widehat{B}$,
which is the projection on $B$, is a meaningful proposition. 
If one has
performed a measurement whose result indicates $A$ and, subsequently,
one performs a measurement whose result indicates $B$, the knowledge that
one possesses about the system is encapsulated 
in the subspace $\widehat{B}(A)$.

At this point, a very fundamental remark kicks in. 
The projection $\widehat{B}(A)$
of a closed subspace $A$ on a closed subspace $B$ is a subspace but is 
not always closed. I am indebted to Semyon Alesker, Joseph Bernstein 
and Vitali Milman for enlightening me and providing me with an explicit 
counter-example.
The counter-example is based on an unbounded operator whose graph is closed.
By a result of Banach (1932) no such operator can be defined on the whole
space, and one must build one such operator defined on only part of the
space.
   
There is no way, then, we can consider an arbitrary Hilbert space \cH\ and
the family of all closed subspaces of \cH. We could decide to consider
only those Hilbert spaces \cH\ for which the set of all closed subspaces is
closed under projections, but there is absolutely no reason to stick
to the idea, discussed critically by Birkhoff and von Neumann, that we
should consider all closed subspaces of \cH. It seems much more natural
not to put restrictions on \cH\ but to consider only families of 
closed subspaces that are closed under projections. This is what will
be done in Section~\ref{sec:semantics}.

\section{Syntax} \label{sec:syntax}
A syntax for denoting {\em measurements} and propositions to talk about them
will be described now. Terms denote measurements.

\begin{definition} \label{def:terms}
Let $V$ be a denumerable set (of atomic terms).
The set of {\em quantic terms} over $V$ will be denoted by $QTerms(V)$
and is defined inductively by:
\begin{enumerate}
\item an element of $V$ (an atomic term) is a quantic term,
\item $1$ is a quantic term,
\item if $x$ is a quantic term, then $\neg x$ is a
quantic term,
\item \label{*_rule} if $x$ and $y$ are quantic terms, then 
\mbox{$x * y$} is a quantic term, and
\item these are the only quantic terms.
\end{enumerate}
\end{definition}
We shall write quantic terms using parentheses when useful and
assuming that $\neg$ has precedence over $*$.

One could consider a more extreme minimalistic approach based on the 
following remark.
If one reflects on the two expressions \mbox{$(x * y) * z$} and
\mbox{$x * (y * z)$}, one notices that the former has an immediate experimental
interpretation: the system may result from a measurement $x$ followed by $y$ 
followed by $z$. The latter expression does not present such a natural 
interpretation. Its meaning is that the system may result from a measurement
of $x$ and then a measurement that it could have been the case that $y$ and 
then $z$ were measured: a quite unnatural proposition to make, since it is not
clear how one could measure that the system could have been in a state 
satisfying $y$ and then $z$ without measuring first $y$ and then $z$.
Therefore, one could have restricted the rule~\ref{*_rule}) above to:
if $x$ is a quantic term and $y$ is a literal (i.e., atomic term or negation
of an atomic term), then \mbox{$x * y$} is a quantic term. This interesting
possibility would probably be best treated in the framework of a calculus
of sequents, and is left for future work.

Propositions talk about terms.
\begin{definition} \label{def:atprop}
A {\em simple} quantic proposition on $V$ is a pair of elements of 
$QTerms(V)$, written \mbox{$x = y$} for \mbox{$x, y \in QTerms(V)$}.
The {\em conditional} quantic propositions on $V$ are defined 
in the following way:
\begin{enumerate}
\item a simple quantic proposition is a conditional quantic proposition,
\item if \mbox{$x = y$} is a simple quantic proposition and $P$ is a 
{\em conditional} quantic proposition then
{\em {\bf if} x = y {\bf then} P} is a conditional quantic proposition, and
\item these are the only conditional quantic propositions.
\end{enumerate}
\end{definition}
\begin{notation}
The proposition {\em {\bf if} w = x {\bf then} {\bf if} y = z {\bf then} P}
will be denoted: {\em {\bf if} w = x {\bf and} y = z {\bf then} P}.
The simple proposition \mbox{$x * y = x$} will denoted \mbox{$x \leq y$}.
\end{notation}

In Section~\ref{sec:semantics} we shall propose a semantics for the calculus
of conditional quantic propositions, based on the geometry of Hilbert spaces.

\section{Semantics} \label{sec:semantics}
We shall formally define the families of closed subspaces we are interested
in.
\begin{definition}
Let \cH\ be a Hilbert space and $M$ be a family of closed subspaces of \cH.
The family $M$ is said to be {\em a P-family} iff
\begin{itemize}
\item \mbox{$\cH \in M$},
\item for any \mbox{$A \in M$}, \mbox{$A^{\perp} \in M$},
\item for any \mbox{$A, B \in M$}, \mbox{$\widehat{B}(A) \in M$}.
\end{itemize}
\end{definition}
Set-theorists: note that we use the term {\em family} 
only for convenience since the families considered are sets.
Note that, as mentioned in Section~\ref{sec:intro}, 
the projection $\widehat{B}(A)$
is not always a closed subspace: $M$ is a P-family only if such projections
amongst members of the family are closed.
There are many examples of P-families. For example, the set of all closed
subspaces of a finite-dimensional Hilbert space is a P-family.
For any Hilbert space \cH, the family containing two elements: \cH\ and
the null subspace is a P-family.

An interpretation $f$ of $QTerms(V)$ into a P-family $M$ of \cH\ 
associates with every quantic term an element of $M$ such that:
\begin{itemize}
\item \mbox{$f(1) = \cH$},
\item \mbox{$f(\neg x) = f(x)^{\perp}$},
\item \mbox{$f(x * y) = \widehat{f(y)}(f(x))$}. 
\end{itemize}

\begin{definition} \label{def:satisfaction}
If $x = y$ is a simple quantic proposition over $V$, and $f$ is an
interpretation of $QTerms(V)$ into a P-family $M$, we shall say that
$x = y$ is {\em satisfied} under $f$ iff \mbox{$f(x) = f(y)$}.
For a conditional quantic proposition {\em {\bf if} x = y {\bf then} P} 
we shall say that it is satisfied under $f$ iff either $P$ is satisfied
under $f$ or $x = y$ is not satisfied under $f$.
A simple (resp. conditional) proposition is {\em valid in a P-family} $M$
iff it is satisfied under any interpretation $f$ into $M$.
A simple (resp. conditional) proposition is {\em Hilbert-valid}
iff it is valid in any P-family.
\end{definition}

The relation $\leq$ defined following Definition~\ref{def:atprop} 
is interpreted
as subset inclusion.
\begin{lemma} \label{le:implication}
Let $f$ be an interpretation of $QTerms(V)$ into a P-family $M$.
The simple proposition \mbox{$x \leq y$} is satisfied under $f$ iff
\mbox{$f(x) \subseteq f(y)$}.
\end{lemma}
\begin{proof}
Let the closed subspaces of \cH, $A$ and $B$ be defined
by: \mbox{$A = f(x)$} and \mbox{$B = f(y)$}.
We see that \mbox{$A \subseteq B$} iff \mbox{$\widehat{B}(A) = A$}
iff \mbox{$f(x * y) = f(x)$}.
\end{proof}

Hilbert Space Quantic Logic is defined to be the set of all Hilbert-valid 
conditional propositions.

\section{Non-Commutative, Non-Associative Boolean algebras} 
\label{sec:ncb-alg}
In this section, an effort is made to try and define the algebraic structures
that can be taken as the essence of Quantum Logic. 
Three principles are guiding us:
\begin{itemize}
\item Language: we are looking for a family of general algebras whose type
consists of two constants, a unary operation and a binary operation.
Clearly other presentations may be considered, in a way that is similar to the
many presentations of Boolean algebras. The only properties 
that we shall consider are properties that can be expressed 
as conditional propositions.
\item Every P-family defines a structure in the family. This is a disputable 
assumption: one may think that not all P-families are meaningful for Quantum
Mechanics and therefore that we may have to consider a subclass of P-families.
In this paper only conditional propositions that are valid amongst all
P-families will be considered.
\item Every Boolean Algebra is an algebra of the family. This assumption is
based on the strong feeling that Quantum Logic should not be seen as 
incompatible with classical logic, as is the case with the currently prevailing
view of Quantum Logic, as attested by the results of Kochen and Specker, but
that classical logic should be a special case of Quantum Logic. More precisely,
classical logic is Commutative Quantum Logic (when for every $x$, $y$,
\mbox{$x * y = y * x$}). 
\end{itemize}

We consider structures \mbox{$\langle M , 0 , 1 , \neg , * \rangle$} where
$M$ is a non-empty set, $0$ and $1$ are elements of $M$, $\neg$ is a unary
function \mbox{$M \longrightarrow M$} and $*$ is a binary function
\mbox{$M \times M \longrightarrow M$}.
\begin{definition} \label{def:NCNAB}
A structure \mbox{$\langle M , 0 , 1 , \neg , * \rangle$} is a
{\em non-commutative, non-associative Boolean algebra} 
(NCNAB-algebra)
iff it satisfies, for all interpretations of atomic terms in $M$, 
all conditional quantic propositions valid in Hilbert Space Quantum Logic.
\end{definition}
Note that Definition~\ref{def:NCNAB} does not require that $0$ be different 
from $1$.

It would be nice to be able to present now a list of conditional quantic 
propositions valid in Hilbert Space Quantum logic and show that any structure
satisfying those propositions is (isomorphic to) an NCNAB-algebra.
This paper does not provide such a completeness result.
  
We shall present a number of conditional quantic propositions that are
valid in Hilbert Space Quantum Logic and prove interesting properties
for all structures that satisfy those properties, and therefore also for any 
NCNAB-algebra.
No claim is made about the completeness of the list, and no
claim is made about the independence of the properties listed in the sequel.

In Section~\ref{sec:star}, we shall present propositions that do not contain 
$\neg$. 
A first result claims that they are valid in Hilbert Space Quantum Logic.
Its proof is postponed to Section~\ref{sec:orthomod}.
A second result shows that in any structure satisfying those propositions,
the relation $\leq$ is a partial order.
In Section~\ref{sec:negation}, we shall present propositions 
that deal with $\neg$, claim that they are valid in 
Hilbert Space Quantum Logic (proof postponed)
and show that any structure that satisfies those propositions 
and those of Section~\ref{sec:star} and is commutative (or associative,
or monotonic) is a Boolean algebra.
In Section~\ref{sec:misc} we shall present valid propositions which, 
at this stage, cannot be proven to follow from the propositions of 
Sections~\ref{sec:star} and~\ref{sec:negation}. 
The reader should notice that all the propositions presented
below have a natural flavor and represent ways of proving properties of quantic
systems.

\subsection{Properties of {\em and then}} \label{sec:star}
Our first set of propositions deal with $*$ only.
We shall say that
{\em $x$ and $y$ commute} if
\mbox{$x * y =$} \mbox{$y * x$}.
\begin{theorem} \label{the:star}
The following conditional quantic propositions are valid in
Hilbert Space Quantum Logic.
\begin{enumerate}
\item \label{global_commutativity} {\bf Global Cautious Commutativity}
if \mbox{$x * y \leq x$} then \mbox{$x * y =$} \mbox{$y * x$},
\item \label{associativity} {\bf Cautious Associativity}
if \mbox{$x * y =$} \mbox{$y * x$}, then, for any \mbox{$z \in M$},
\mbox{$z * (x * y) =$} \mbox{$(z * x) * y$},
\item \label{commutativity}  
{\bf Local Cautious Commutativity} if 
\mbox{$(z * x) * y \leq x$} and \mbox{$(z * y) * x \leq y$}, then 
\mbox{$(z * x) * y =$} \mbox{$(z * y) * x$},
\item \label{zero} {\bf Z} \mbox{$0 * x = 0 = x * 0$},
\item \label{one} {\bf N} \mbox{$1 * x = x = x * 1$},
\item \label{left_monotony} {\bf Left Monotony} if \mbox{$x \leq y$},
then, \mbox{$x * z \leq y * z$}.
\end{enumerate}
\end{theorem}

Remarks:
\begin{itemize}
\item the binary operation $*$ is not assumed to be associative or commutative.
\item Taking $M$ to be a Boolean algebra, $0$ to be the bottom element, 
$1$ the top element, $\neg$ to be complementation and $*$ to be 
greatest lower bound, 
one obtains a model of all of the properties above,
in which $*$ is associative and commutative, as well as a model of the 
properties of Theorems~\ref{the:negation} and~\ref{the:misc}.
\item {\bf Global Cautious Commutativity} 
(GCC) is a weak commutativity property,
it claims that, under certain circumstances, $*$ is commutative.
The commutativity property asserted \mbox{$x * y = y * x$} represents
a global commutation property: $x$ and $y$ commute in any context.
Commutation in a specific context $z$, a local commutation property,
is expressed as \mbox{$(z * x) * y = (z * y) * x$} and appears in
the property of {\bf Local Cautious Commutativity} (LCC) below.
Theorem~\ref{the:partial_order}, item~\ref{C<=}) 
shows that two propositions that 
commute globally, commute locally in any context.
\item {\bf Cautious Associativity} (CA) is a weak associativity property:
under certain circumstances, i.e., if $x$ and $y$ commute, we have 
associativity for $z$, $x$ and $y$.
\item LCC is a weak commutativity property, it claims that, 
under certain circumstances, propositions $x$
and $y$ commute {\em locally}, i.e., in the context of $z$.
\item Z expresses the fact that $0$ is a zero for the operation $*$.
\item N expresses the fact that $1$ is a neutral element for 
the operation $*$.
\item {\bf Left Monotony} (LM) expresses the fact that the operation $*$
is monotone, with respect to $\leq$, in its left argument.
A symmetric property of right monotony would imply commutativity since
\mbox{$x \leq 1$} would imply \mbox{$y * x \leq$} \mbox{$y * 1 = y$} and
GCC would then imply \mbox{$x * y = y * x$}.
\end{itemize}
\begin{proof}
One could prove directly, without much difficulty, that the properties
of Theorem~\ref{the:star} are valid in Hilbert Space Logic.
Since a stronger result, validity in Orthomodular Logic, will
be proved in Theorem~\ref{the:ortho}, we postpone the proof.
\end{proof}

We may now prove that any structure satisfying the properties of
Theorem~\ref{the:star} has many interesting properties.
\begin{theorem} \label{the:partial_order}
The following properties hold in any structure that satisfies 
the properties GLC, CA, 
LCC, Z, N, and LM of Theorem~\ref{the:star}:
\begin{enumerate}
\item \label{zero_one} \mbox{$0 \leq x \leq 1$},
\item \label{LI} \mbox{$x * y \leq y$},
\item \label{reflexivity} \mbox{$x \leq x$}, i.e., the relation $\leq$ is
reflexive, i.e., \mbox{$x * x = x$},
\item \label{<=comm} if \mbox{$x \leq y$} 
then $x$ and $y$ commute,
\item \label{antisymmetric} the relation $\leq$ is antisymmetric,
\item \label{transitive} the relation $\leq$ is transitive,
\item \label{partial_order} the relation $\leq$ is a partial order,
\item \label{full_AC} if \mbox{$x * y =$} \mbox{$y * x$}, then, 
for any \mbox{$z \in M$} we have:
\mbox{$z * (y * x) =$} \mbox{$z * (x * y) =$} \mbox{$(z * x) * y = $}
\mbox{$(z * y) * x$},
\item \label{C<=} if \mbox{$x * y = y * x$}, then, for any \mbox{$z \in M$}
we have:
\mbox{$(z * x) * y \leq x$} (and \mbox{$(z * y) * x \leq y$}),
\item \label{p<=} if \mbox{$x \leq y$}, then for any \mbox{$z \in M$}:
\mbox{$z * x = (z * y) * x$}.
\end{enumerate}
\end{theorem}
\begin{proof}
\begin{enumerate}
\item By Z and N.
\item By~\ref{zero_one}) above, 
\mbox{$x \leq 1$}. By LM, \mbox{$x * y \leq 1 * y$}.
By N, \mbox{$x * y \leq y$}.
\item By~\ref{LI}) above, \mbox{$1 * x \leq x$} and now, by N, 
\mbox{$x \leq x$}.
\item If \mbox{$x * y = x$}, by~\ref{reflexivity}) of this Lemma, 
\mbox{$x * y \leq x$} and, by (GCC), $x$ and $y$ commute.
\item Assume \mbox{$x \leq y$} and \mbox{$y \leq x$}.
By~\ref{<=comm}) above, $x$ and $y$ commute.
But \mbox{$x * y = x$} and \mbox{$y * x = y$}.
We conclude that \mbox{$x = y$}.
\item Assume \mbox{$x \leq y$} and \mbox{$y \leq z$}.
We have \mbox{$x =$}
\mbox{$x * y =$} \mbox{$x * (y * z)$}.
But, by \ref{<=comm}) above, $y$ and $z$ commute and, therefore, by CA
we have \mbox{$x * (y * z) =$} \mbox{$(x * y) * z =$} \mbox{$x * z$}.
\item Obvious from the above.
\item From the assumption: \mbox{$z * (y * x) =$} \mbox{$z * (x * y)$}.
By CA \mbox{$(z * x) * y =$} \mbox{$z * (x * y)$} and also
\mbox{$(z * y) * x =$} \mbox{$z * (y * x)$}.
\item
By~\ref{full_AC}) above, and then \ref{LI}) above, \mbox{$(z * x) * y =$}
\mbox{$z * (y * x) \leq$} \mbox{$y * x \leq x$}. 
By \ref{transitive}) above, we conclude that \mbox{$(z * x) * y \leq x$}.
\item By~\ref{<=comm}) $x$ and $y$ commute and by~\ref{full_AC}) 
\mbox{$(z * y) * x =$} \mbox{$z * (x * y) =$} \mbox{$z * x$}.
\end{enumerate}
\end{proof}

\subsection{Properties of negation} \label{sec:negation}
We shall now deal with properties that involve both $*$ and $\neg$.
We shall write \mbox{$x \perp y$} for \mbox{$x * y = 0$}.
\begin{theorem} \label{the:negation}
The following conditional quantic propositions are valid in
Hilbert Space Quantum Logic.
\begin{enumerate}
\item \label{negation1} {\bf NP} \mbox{$x * \neg x = \neg x * x = 0$},
i.e., \mbox{$x \perp \neg x$} and \mbox{$\neg x \perp x$},
\item \label{negation2} {\bf RNL} if \mbox{$x * z \leq y$} and
\mbox{$x * \neg z \leq y$}, then
\mbox{$x \leq y$}.
\end{enumerate}
\end{theorem}

Remarks:
\begin{itemize}
\item NP, and RNL may be considered to be the proof rules that
define negation.
NP parallels a left introduction rule. 
RNL is a non-commutative left elimination rule.
\item The property LNL, dual to RNL and expressed:
if \mbox{$z * x \leq y$} and
\mbox{$\neg z * x \leq y$}, then 
\mbox{$x \leq y$} is also valid in Hilbert Space Quantum Logic.
It will be described and discussed in Section~\ref{sec:misc}.
\item The properties RNL and LNL are an important novelty of this paper.
All the properties of Theorems~\ref{the:star} and~\ref{the:negation}, 
except RNL, 
are satisfied in Hilbert space when $*$ is interpreted as intersection
and $\neg$ as orthogonal complement, the interpretation 
proposed by~\cite{BirkvonNeu:36}.
Neither RNL nor its dual LNL are satisfied in this interpretation. 
Both are very natural rules that express a very basic
rule of reasoning, reasoning by cases: to prove $\alpha$ it is enough to 
prove that $\alpha$ holds if $\beta$ holds {\em and} that $\alpha$ holds
if $\neg \beta$ holds. 
Such reasoning by cases is valid in classical logic. It is also valid in many
(preferential) non-monotonic logics~\cite{KLMAI:89}.
It is also used in Quantum Physics.
The following presents a use of RNL.
To prove that a system prepared in a certain way has a certain 
quantic property, it is enough to show that, after some measurement, all
possible resulting systems have the property. 
Suppose, for example, that one prepares many copies of a quantic system 
and then measures, on each copy, its spin along some direction $d'$.
One finds many possible values for the spin along the direction $d'$.
If, then, on each of the resulting systems (with different values for the
spin along $d'$) one measures the value $0$ for the spin along a direction $d$,
this is a proof that the original system (before measuring along $d'$) had
a zero spin along $d$.
Such a proof-rule seems to be crucially needed because, even if one measures
the spin along $d$ immediately (without measuring first along $d'$) one cannot,
in effect, exclude the possibility that some interaction between the system 
and its environment occured, resulting in some unknown measurement.
\end{itemize}

\begin{proof}
As for Theorem~\ref{the:star}, the proof is postponed to 
Theorem~\ref{the:ortho}.
\end{proof}

A series of theorems will now describe properties of all structures
satisfying the properties above.
\begin{theorem} \label{thele:negation}
The following properties hold in any structure that satisfies 
the properties GLC, CA, LCC, Z, N, and LM of Theorem~\ref{the:star} and
the properties NP and RNL of Theorem~\ref{the:negation}.
\begin{enumerate}
\item \label{doubleneg} \mbox{$\neg (\neg x) = x$},
\item \label{zero_new} \mbox{$0 = \neg 1$} and \mbox{$1 = \neg 0$},
\item \label{perpsymmetric} the relation $\perp$ is symmetric,
\item \label{<=_perp} \mbox{$x \leq y$} iff \mbox{$x \perp \neg y$},
\item \label{antimon} \mbox{$x \leq y$} iff 
\mbox{$\neg y \leq \neg x$},
\item \label{perp<=} if \mbox{$x \leq y$} and \mbox{$y \perp z$},
then \mbox{$x \perp z$},
\item \label{bottom} if \mbox{$y \leq x$} and 
\mbox{$y \leq \neg x$} then \mbox{$y = 0$},
\item \label{top} if \mbox{$x \leq y$} and 
\mbox{$\neg x \leq y$}, then \mbox{$y = 1$},
\item \label{and}
if \mbox{$x \leq y$} and \mbox{$x \leq z$}, then
\mbox{$x \leq y * z$},
\end{enumerate}
\end{theorem}
\begin{proof}
\begin{enumerate}
%doubleneg
\item By Theorem~\ref{the:partial_order}, item~\ref{LI})
\mbox{$\neg \neg x * x \leq x$}.
By NP and Theorem~\ref{the:partial_order}, item~\ref{zero_one}) 
\mbox{$\neg \neg x * \neg x = 0 \leq x$}.
We conclude, by RNL, that \mbox{$\neg \neg x \leq x$}.
Similarly we can show that \mbox{$x \leq \neg \neg x$}.
We conclude, by Theorem~\ref{the:partial_order}, that
\mbox{$x = \neg \neg x$}.
%zero_new
\item By NP \mbox{$\neg 1 * 1 = 0$}. By N \mbox{$\neg 1 * 1 = \neg 1$}.
Therefore \mbox{$\neg 1 = 0$} and, by~\ref{doubleneg}) above, we have
\mbox{$\neg 1 = \neg \neg 0 = 0$}.
%perpsymmetric
\item if \mbox{$x * y = 0$}, then, by Theorem~\ref{the:partial_order}, 
item~\ref{zero_one}), \mbox{$x * y \leq x$} and, by 
Theorem~\ref{the:partial_order}, item~\ref{<=comm}) 
$x$ and $y$ commute and therefore
\mbox{$y * x = 0$}.
%<=_perp
\item If \mbox{$x \leq y$}, we have \mbox{$x * \neg y =$}
\mbox{$(x * y) * \neg y$}. But, by NP, $y$ and $\neg y$ commute
and therefore, by CA and then NP and Z, 
\mbox{$x * \neg y =$} \mbox{$x * (y * \neg y) =$} \mbox{$x * 0 = 0$}.

If \mbox{$x * \neg y = 0$}, then \mbox{$x * \neg y \leq y$} 
by Theorem~\ref{the:partial_order}, item~\ref{zero_one}).
But \mbox{$x * y \leq y$} by Theorem~\ref{the:partial_order}, item~\ref{LI}). 
We conclude, by RNL, that \mbox{$x \leq y$}.
%antimon
\item \mbox{$x \leq y$} iff, by~\ref{<=_perp}), \mbox{$x \perp \neg y$}, 
iff, by~\ref{perpsymmetric}), \mbox{$\neg y \perp x$} iff, 
by~\ref{doubleneg}), \mbox{$\neg y \perp \neg \neg x$} iff, 
by~\ref{<=_perp}), \mbox{$\neg y \leq \neg x$}. 
\item If \mbox{$y \perp z$}, we have, by~\ref{doubleneg}),
\mbox{$y \perp \neg \neg z$}
and, by~\ref{perp<=}) \mbox{$y \leq \neg z$}. By transitivity of $\leq$ we have
\mbox{$x \leq \neg z$} and therefore
\mbox{$x \perp \neg \neg z$} and \mbox{$x \perp z$}.
\item \mbox{$y \leq x$} implies \mbox{$y * \neg x \leq 0$}.
\mbox{$y \leq \neg x$} implies \mbox{$y * \neg \neg x = 0$}
and \mbox{$y * x \leq 0$}. By RNL, then, \mbox{$y \leq 0$} and since
\mbox{$0 \leq y$}, \mbox{$y = 0$} by Theorem~\ref{the:partial_order}, 
item~\ref{antisymmetric}).
\item Assume \mbox{$x \leq y$} and \mbox{$\neg x \leq y$}.
By~\ref{antimon}) above we have \mbox{$\neg y \leq \neg x$} and 
\mbox{$\neg y \leq$} \mbox{$\neg \neg x =$} \mbox{$x$} and, 
by~\ref{bottom}), we have
\mbox{$\neg y = 0$}, therefore 
\mbox{$y =$} \mbox{$\neg \neg y =$}
\mbox{$\neg 0 = 1$} by~\ref{zero_new}).
\item Assume \mbox{$x \leq y$} and \mbox{$x \leq z$}. By LM,
\mbox{$x * z \leq$} \mbox{$y * z$}. But, by~\ref{<=_perp}) 
\mbox{$x * \neg z =$} \mbox{$0 \leq$} \mbox{$y * z$}.
By RNL, then, \mbox{$x \leq$} \mbox{$y * z$}.
\end{enumerate}
\end{proof}

The next lemma deals with commuting propositions.
\begin{lemma} \label{le:commuting_prop}
\begin{enumerate}
In any structure that satisfies the properties of Theorems~\ref{the:star} 
and~\ref{the:negation}:
\item \label{comm_*}
if all three propositions $x$, $y$ and $z$ commute pairwise, then $x$ commutes
with \mbox{$y * z$},
\item \label{comm_neg} if $x$ commutes with $y$, then $x$ commutes 
with $\neg y$,
\item \label{glb_lub} if $x$ and $y$ commute, then $x * y$ is their greatest 
lower bound and \mbox{$\neg(\neg x * \neg y)$} their least upper bound,
\item \label{prelim_Robbins} if $x$ and $y$ commute, then
\mbox{$\neg (x * y) * y \leq$} \mbox{$\neg x$},
\item \label{Robbins} {\bf Robbins equation}
if $x$ and $y$ commute then 
\mbox{$x =$} \mbox{$\neg ( \neg (x * y) * \neg (x * \neg y))$},
\item \label{orthomod} {\bf Orthomodularity} if \mbox{$x \leq y$},
then $y$ is the least upper bound of $x$ and $\neg x * y$.
\end{enumerate}
\end{lemma}
\begin{proof}
\begin{enumerate}
\item By CA \mbox{$x * (y * z) = (x * y) * z$} since $y$ and $z$ commute.
Since $x$ and $y$ commute \mbox{$(x * y) * z = (y * x) * z$}.
But $x$ and $z$ commute and, by Theorem~\ref{the:partial_order},
item~\ref{full_AC})
\mbox{$(y * x) * z =$} \mbox{$(y * z) * x$}.
\item Assume $x$ and $y$ commute. We have, by Z, NP and 
Theorem~\ref{the:partial_order}, item~\ref{full_AC}):
\[
0 = 0 * y = (\neg x * x) * y = (\neg x * y) * x.
\]
Therefore \mbox{$\neg x * y \perp x$}, \mbox{$\neg x * y \perp \neg \neg x$},
\mbox{$\neg x * y \leq \neg x$} and, by GCC, $\neg x$ and $y$ commute.
\item For arbitrary $x$ and $y$, \mbox{$x * y \leq y$} 
by Theorem~\ref{the:partial_order}, item~\ref{LI});
also $x * y$ is greater or equal to any lower bound of $x$ and
$y$, by Theorem~\ref{thele:negation}, item~\ref{and}).
The fact that $x$ and $y$ commute gives us the last property needed:
\mbox{$x * y =$} \mbox{$y * x \leq x$} by Theorem~\ref{the:partial_order},
item~\ref{LI}).

By~\ref{comm_neg} $\neg x$ and $\neg y$ commute. Therefore 
$\neg x * \neg y$ is the greatest lower bound of $\neg x$ and $\neg y$.
By Theorem~\ref{thele:negation}, item~\ref{antimon}), 
\mbox{$\neg (\neg x * \neg y)$}
is therefore the least upper bound of $\neg \neg x$ and $\neg \neg y$.
\item This is property (4) of Finch~\cite{Finch_lattice:69}, for the special
case $x$ and $y$ commute. The claim holds without this assumption,
see Theorem~\ref{the:misc}.
Assume $x$ and $y$ commute.
By Theorem~\ref{the:partial_order}, item~\ref{LI}), we have
\mbox{$(\neg (x * y) * y) * \neg x \leq$} \mbox{$\neg x$}.
But, by CA, we have: 
\[
(\neg (x * y) * y) * x = \neg (x * y) * (y * x) = 
\neg (x * y) * (x * y) = 0 \leq \neg x.
\]
By RNL we conclude that
\mbox{$\neg (x * y) * y \leq$} \mbox{$\neg x$}.
\item It is enough to prove that, if $x$ and $y$ commute
\mbox{$ \neg x =$} \mbox{$\neg (x * y) * \neg (x * \neg y)$}.
We have: \mbox{$x * y =$} \mbox{$y * x \leq$} \mbox{$x$} 
by Theorem~\ref{the:partial_order}, item~\ref{LI}) and therefore, 
by 
Theorem~\ref{thele:negation}, item~\ref{antimon}) \mbox{$\neg x \leq$}
\mbox{$\neg (x * y)$}.
By~\ref{comm_neg}) above, $x$ commutes with $\neg y$ and
\mbox{$x * \neg y =$} \mbox{$\neg y * x \leq$} \mbox{$x$} 
by Theorem~\ref{the:partial_order}, item~\ref{LI}) and therefore, 
by 
Theorem~\ref{thele:negation}, item~\ref{antimon}) 
\mbox{$\neg x \leq$} \mbox{$\neg (x * \neg y)$}.
By Theorem~\ref{thele:negation}, item~\ref{and}), we have
\mbox{$\neg x \leq$} \mbox{$\neg (x * y) * \neg (x * \neg y)$}.

Consider, now that, by~\ref{comm_*}) and~\ref{comm_neg}) just above $x$,
\mbox{$\neg ( x * y)$} and \mbox{$\neg (x * \neg y)$} commute pairwise.
By CA, then, we have
\[  
(\neg (x * y) * \neg (x * \neg y)) * x = 
\neg (x * y) * (\neg (x * \neg y) * x)) \leq \neg (x * \neg y) * x \leq y
\] 
by~\ref{prelim_Robbins}) above,
but we also have
\[
(\neg (x * y) * \neg (x * \neg y)) * x =
\neg ( x * \neg y) * (\neg (x * y) * x)
\leq
\neg (x * y) * x) \leq \neg y
\]
by~\ref{prelim_Robbins}) above.
By Theorem~\ref{thele:negation}, item~\ref{bottom}),
\[
(\neg (x * y) * \neg (x * \neg y)) * x =
0.
\]
Now, by Theorem~\ref{thele:negation}, item~\ref{<=_perp}),
\mbox{$(\neg (x * y) * \neg (x * \neg y)) \leq \neg x$}.
\item If \mbox{$x \leq y$} then clearly $y$ is an upper bound
for $x$ and for $(\neg x) * y$ by Theorem~\ref{the:partial_order}, item~\ref{LI}).
Suppose now that \mbox{$x \leq z$} and
\mbox{$(\neg x) * y \leq$} \mbox{$z$}.
Since \mbox{$x \leq y$}, $x$ and $y$ commute, and, by just above,
\mbox{$y =$} \mbox{$\neg ( \neg ( y * x) * \neg (y * \neg x))$}.
Therefore \mbox{$y =$}
\mbox{$\neg ( \neg x * \neg (y * \neg x))$}.
By~\ref{glb_lub} above, $y$ is the least upper bound of $x$ and 
\mbox{$y * \neg x =$} \mbox{$\neg x * y$}.
\end{enumerate}
\end{proof}

\begin{definition} \label{def:commutative}
A structure is {\em commutative} iff for any \mbox{$x, y \in M$},
\mbox{$x * y =$} \mbox{$y * x$}.
A structure is {\em associative} iff for any \mbox{$x, y, z \in M$},
\mbox{$(x * y) * z =$} \mbox{$x * (y * z)$}.
A structure is {\em monotone} iff for any \mbox{$x, y \in M$},
\mbox{$x * y \leq x$}.
\end{definition}

\begin{theorem} \label{the:comm_ass}
For a structure \mbox{$A = \langle M, 0, 1, \neg, *\rangle$} satisfying
the properties of Theorems~\ref{the:star} and~\ref{the:negation} 
the following propositions are equivalent:
\begin{enumerate}
\item $A$ is associative,
\item $A$ is monotone,
\item $A$ is commutative,
\item $A$ is a Boolean algebra.
\end{enumerate}
\end{theorem}
The failure of monotonicity is a hallmark of the approach to Quantum Logic 
taken in~\cite{EngGabbay:Quantum}. 
Theorem~\ref{the:comm_ass} shows that this failure is inherently
linked to the failure of associativity and commutativity.
It was the feeling of many that, since the hallmark of Quantum Mechanics,
as opposed to Classical Mechanics, is the non-commutativity of operators, 
Quantum Logic should, in some way, be non-commutative.
Theorem~\ref{the:comm_ass} shows why it also has to be non-associative,
a property that is more surprising.
\begin{proof}
%Associative implies Monotone implies Commutative implies Boolean implies Associative.
Assume $A$ is associative. Consider arbitrary elements $x$ and $y$. 
We shall show that \mbox{$x * y \leq x$}.
By associativity: \mbox{$(x * y) * \neg x =$} \mbox{$x * (y * \neg x)$}.
But, by Theorem~\ref{the:partial_order}, item~\ref{LI}),
\mbox{$y * \neg x \leq \neg x$} and, 
by Theorem~\ref{thele:negation}, item~\ref{perp<=})
and \ref{doubleneg}): \mbox{$y * \neg x \perp \neg \neg x = x$}.
Therefore, by Theorem~\ref{thele:negation}, item~\ref{perpsymmetric})
we have \mbox{$x * (y * \neg x) = 0$} and
\mbox{$(x * y) * \neg x = 0$}, \mbox{$x * y \perp \neg x$} and,
by Theorem~\ref{thele:negation}, item~\ref{<=_perp}) \mbox{$x * y \leq x$}.

If $A$ is monotone, then, by GCC, it is commutative.

Assuming $A$ is commutative, we could use any of many different 
characterizations of Boolean algebras to show that it is a Boolean algebra.
We shall use the one conjectured by Robbins. McCune~\cite{McCune:Robbins}
proved Robbins conjecture: any structure in which $*$ is associative,
commutative and satisfies the Robbins equation, 
for any elements $x$ and $y$:
\[
\neg ( \neg x * y ) * \neg ( \neg x * \neg y ) = x,
\]
is a Boolean algebra.
The operation $*$ is commutative by assumption. 
It is associative by CA.
It satisfies the Robbins equation by 
Lemma~\ref{le:commuting_prop}, item~\ref{Robbins}).

A Boolean algebra is associative.
\end{proof}

\begin{definition} \label{def:subalgebra}
Let $M$ be any NCNAB-algebra and let \mbox{$X \subseteq M$} be a set
of propositions of $M$. The sub-algebra generated by $X$, $M(X)$ is
the smallest sub-algebra of $M$ containing $X$.
\end{definition}
Note that $M(X)$ is an NCNAB-algebra since the intersection of a family of
NCNAB-algebras is an NCNAB-algebra due to the conditional-equational form of
the properties defining an NCNAB-algebra.

\begin{lemma} \label{le:sub-alg}
Let $M$ be any NCNAB-algebra and let \mbox{$X \subseteq M$} be a set
of pairwise commuting propositions: i.e., for any \mbox{$x, y \in X$}
\mbox{$x * y = y * x$}, then the sub-algebra of $M$ generated by $X$, $M(X)$
is a commutative NCNAB-algebra.
\end{lemma}
\begin{proof}
By Lemma~\ref{le:commuting_prop}, items~\ref{comm_*}) and \ref{comm_neg}). 
\end{proof}

\subsection{Additional propositions valid in Hilbert Space Quantum Logic}
\label{sec:misc}
Some additional propositions that are valid in Hilbert Space Quantum Logic
will be presented here. The question whether
these properties follow from those of Theorems~\ref{the:star} 
and~\ref{the:negation} is still open.
\begin{theorem} \label{the:misc}
The following properties hold in any NCNAB-algebra.
\begin{enumerate}
\item \label{negation3} {\bf LNL} if \mbox{$z * x \leq y$} and
\mbox{$\neg z * x \leq y$}, then \mbox{$x \leq y$},
\item \label{NN} {\bf NN} if \mbox{$x \leq y$} and 
\mbox{$x * \neg z \leq y$}, then \mbox{$x * z \leq y$},
\item \label{F4} {\bf F4} \mbox{$y * (x * y)' \leq x'$}.
\end{enumerate}
\end{theorem}
LNL is the dual of RNL. NN is a paradoxical rule of proof: to prove $y$ after
one measures $x$ and $z$, it is enough to prove .
NN is a rule of cautious monotony and the converse of RNL.
F4 is not easily interpreted in terms of quantic measurements.
F4 is property (4) of Finch~\cite{Finch_lattice:69}. A special case was
proved in Lemma~\ref{le:commuting_prop}, item~\ref{prelim_Robbins}.
\begin{proof}
The proof is postponed to Theorem~\ref{the:ortho}.
\end{proof}
\begin{lemma} \label{le:FN}
In any structure that satisfies the properties of Theorems~\ref{the:star} 
and~\ref{the:negation} and F4, we have
\mbox{$x * (x * y)' \leq y'$}.
\end{lemma}
\begin{proof}
Since \mbox{$x * y \leq y$}, $y$ and \mbox{$(x * y)'$} commute.
Therefore, \mbox{$(x * (x * y)') * y =$}
\mbox{$(x * y) * (x * y)' = 0 \leq y'$}.
But \mbox{$(x * (x * y)') * y' \leq y'$} and we conclude that
\mbox{$x * (x * y)' \leq y'$}.
\end{proof}

\section{Orthomodular and Modular Quantum Logic} \label{sec:orthomod}
A different, weaker, semantics, based on orthocomplemented lattices
may be considered. It was proposed by Finch in~\cite{Finch_lattice:69}.

An interpretation $f$ of $QTerms(V)$ into an orthocomplemented lattice
\mbox{$\langle X , \bot , \top, {\ }' , \leq \rangle$}
associates with every quantic term an element of $M$ such that:
\begin{itemize}
\item \mbox{$f(1) = \top$},
\item \mbox{$f(\neg x) = f(x)'$},
\item \mbox{$f(x * y) = (f(x) \vee f(y)') \wedge f(y)$}. 
\end{itemize}
Quantic propositions are given the obvious interpretation. Validity is defined
as usual, for diferent families of orthocomplemented lattices: orthomodular,
modular, and Boolean algebras.
Orthomodular (resp. modular, Boolean) Quantum Logic is the set of all 
conditional propositions valid in orthomodular (resp. modular, Boolean) 
lattices.
It is easy to see that in Boolean lattices, one has: 
\mbox{$x * y = x \wedge y$} and therefore Boolean Quantum Logic is classical
logic. But even in modular lattices $*$ is different from $\wedge$: consider
the modular lattice of all subspaces of a Hilbert space. 

Let us now sort out the relations between all those logics we considered:
Hilbert Space Quantum Logic (HSQL), Orthomodular Quantum Logic (OQL),
Modular Quantum Logic (MQL) and Boolean Logic (BL).
\begin{theorem} \label{the:hierarchy}
\[
OQL \subseteq MQL \subseteq HSQL \subset BL.
\]
\end{theorem}
The rightmost inclusion is strict. It is not known whether OQL and HSQL
are different.

In~\cite{BirkvonNeu:36}, Birkhoff and von Neumann proposed {\em modular} 
lattices as the structure of Quantum Logic. The research community did not
chose this path and pursued the orthomodular path. Theorem~\ref{the:hierarchy}
shows that, for the limited language considered in this paper, one may go the
modular way.
\begin{proof}
Orthomodular Quantum Logic is a subset of Modular Quantum Logic
since any modular lattice is orthomodular. We do not know whether
the inclusion is strict.
To see that Modular Quantum Logic is a subset of Hilbert Space Quantum Logic
consider that any P-family is part of a modular lattice: the
lattice of all subspaces of \cH. Complementation in the lattice is orthogonal
complementation in Hilbert space. We are left to show that, in a P-family,
the lattice operation defined by Finch is projection.
In other terms, that given any two closed subspaces $A$ and $B$ of the 
P-family, the projection of $A$ on $B$, $\widehat{B}(A)$, is
\mbox{$(A + B^{\bot}) \cap B$}.
\begin{lemma} \label{le:*proj}
Let \cH\ be Hilbert. If $A$ is any (not necessarily closed) linear subspace
of \cH\ and $B$ is any closed subspace of \cH, then 
\mbox{$\widehat{B}(A) =$} \mbox{$(A + B^{\bot}) \cap B$}.
\end{lemma}
\begin{proof}
\mbox{$\vec{u} \in \widehat{B}(A)$} iff 
there is some \mbox{$\vec{v} \in A$}
such that \mbox{$\vec{u} = \widehat{B}(\vec{v})$} iff \mbox{$\vec{u} \in B$} 
and there is some 
\mbox{$\vec{v} \in A$} such that \mbox{$\vec{v} - \vec{u} \perp B$} iff
\mbox{$\vec{u} \in B$} and there are some \mbox{$\vec{v} \in A$} and
\mbox{$\vec{w} \in B^{\perp}$} such that \mbox{$\vec{u} = \vec{v} + \vec{w}$}
iff \mbox{$\vec{u} \in (A + B^{\bot}) \cap B$}.
\end{proof}

It is not known whether Hilbert Space Quantum Logic is
different from Modular Quantum Logic, or even whether it is different
from Orthomodular Logic. The orthoarguesian law of~\cite{Greechie_strong:81} 
that traditionally separates Hilbert space logic from orthomodular logic is 
not obviously expressible in terms of $*$ and $\neg$ only.

Hilbert Space Quantum Logic is a strict subset of Boolean Logic.
Indeed any Boolean Algebra is a field of subsets of some set $X$.
Consider now the Hilbert space whose orthonormal basis is $X$. 
The elements of the field are closed subspaces and they form a P-family.
MQL is therefore a subset of Boolean Logic. It is a strict subset since 
HSQL is not commutative.
\end{proof}

We shall now prove that all the properties of HSQL that were mentioned
in Section~\ref{sec:ncb-alg} are part of OQL, the weakest of our logics,
therefore proving Theorems~\ref{the:star}, \ref{the:negation} 
and~\ref{the:misc}.

Let us assume an orthomodular lattice and define \mbox{$a * b =$}
\mbox{$(a \vee b') \wedge b$}.
First, note that the relation $\leq$ we define in NCNAB-algebras coincides
with the ordering of the lattice. If we use $\leq$ to represent the
order of the lattice: \mbox{$x \leq y$} iff \mbox{$x * y = x$}.
Proof: Assume \mbox{$x \leq y$}, then, by orthomodularity 
\mbox{$x = y \wedge (x \vee y')$}, i.e., \mbox{$x = x * y$}.
Conversely, if \mbox{$x = y \wedge (x \vee y')$}, then \mbox{$x \leq y$}.

\begin{lemma} \label{le:orth_thesis2}
If \mbox{$z * x \leq y$}, then \mbox{$z * x \leq z* (x \wedge y)$}.
\end{lemma}
\begin{proof}
By definition \mbox{$z * x \leq$} \mbox{$z \vee x'$}.
Therefore \mbox{$z * x \leq$}
\mbox{$z \vee x' \vee y' =$}
\mbox{$z \vee (x \wedge y)'$}.
But, by definition \mbox{$z * x \leq x$} and, by assumption,
\mbox{$z * x \leq y$}. We conclude that
\mbox{$z * x \leq$}
\mbox{$(z \vee (x \wedge y)') \wedge x \wedge y =$}
\mbox{$z * (x \wedge y)$}.
\end{proof}
\begin{lemma} \label{le:orth_thesis3}
If \mbox{$x \leq y$}, then for any $z$,
\mbox{$z * x =$} \mbox{$(z * y) * x$}.
\end{lemma}
\begin{proof}
By orthonormality 
\mbox{$z \leq (z \vee y') \wedge y \vee y'$}.
By assumption, \mbox{$y' \leq x'$} and therefore
\mbox{$z \leq (z \vee y') \wedge y \vee x' =$}
\mbox{$(z * y) \vee x'$}.
Therefore \mbox{$z \vee x' \leq$} \mbox{$(z * y) \vee x'$}
and \mbox{$z * x \leq$} \mbox{$(z * y) * x$}.

But \mbox{$y' \leq x'$} implies:
\mbox{$z \vee y' \leq$} \mbox{$z \vee x'$},
and \mbox{$(z \vee y') \wedge y \leq$} \mbox{$z \vee x'$}.
Therefore \mbox{$z * y \leq$} \mbox{$z \vee x'$},
\mbox{$((z * y) \vee x') \wedge x \leq$} \mbox{$(z \vee x') \wedge x$}, i.e.,
\mbox{$(z * y) * x \leq$} \mbox{$z * x$}.
\end{proof}

\begin{lemma} \label{le:orth_thesis4}
If \mbox{$(z * x) * y \leq x$}, then \mbox{$(z * x) * y =$}
\mbox{$z * (x \wedge y)$}.
\end{lemma}
\begin{proof}
Assume \mbox{$(z * x) * y \leq x$}. We have
\mbox{$(z * x) * y \leq$} \mbox{$x \wedge y$} and 
\mbox{$(z * x) * y =$} \mbox{$((z * x) * y) * (x \wedge y)$}.
By Lemma~\ref{le:orth_thesis3}, 
\mbox{$((z * x) * y) * (x \wedge y) =$} \mbox{$(z * x) * (x \wedge y) =$}
\mbox{$z * (x \wedge y)$}.
\end{proof}
The next lemma shows that orthomodular structures satisfy some limited form
of distributivity.
\begin{lemma} \label{le:orth_thesis5}
If \mbox{$z' \leq x$} and \mbox{$z' \leq y$} then
\mbox{$(x \vee y) \wedge z =$} \mbox{$(x \wedge z) \vee (y \wedge z)$}.
Therefore \mbox{$(x \vee y) * z =$} \mbox{$(x * z) \vee (y * z)$}.
\end{lemma}
\begin{proof}
In any lattice and without any assumption
\mbox{$(x \vee y) \wedge z \geq$} \mbox{$(x \wedge z) \vee (y \wedge z)$}.

If \mbox{$z' \leq x$}, we have, by orthomodularity,
\mbox{$x = z' \vee z \wedge x$}. Similarly, \mbox{$z' \leq y$} implies
\mbox{$y = z' \vee z \wedge y$}.
Therefore \mbox{$(x \vee y) \wedge z =$}
\mbox{$(z \wedge x \vee z \wedge y \vee z') \wedge z$}.
But \mbox{$z \wedge x \vee z \wedge y \leq z$} and,
by orthonormality:
\mbox{$z \wedge x \vee z \wedge y =$}
\mbox{$z \wedge (z \wedge x \vee z \wedge y \vee z')$}.
  
The last claim follows trivially.
\end{proof}

\begin{lemma} \label{le:orth_thesis6}
\mbox{$(x \vee y) \wedge (x \vee y') =$}
\mbox{$x \vee (x \vee y') \wedge y$}.
\end{lemma}
\begin{proof}
Without any hypothesis, in any lattice
\mbox{$x \vee (x \vee y') \wedge y \leq$}
\mbox{$(x \vee y) \wedge (x \vee y')$}.

By orthomodularity, it is now enough to show that we have:
\[(x \vee (x \vee y') \wedge y)' \wedge (x \vee y) \wedge (x \vee y') 
= 0,
\] i.e.,
\[x' \wedge (x' \wedge y \vee y') \wedge (x \vee y) \wedge (x \vee y') =
0.
\]
But \mbox{$(x' \wedge y \vee y') \wedge (x \vee y') = y'$} by orthomodularity
since \mbox{$y' \leq x \vee y'$}.
Therefore 
\mbox{$x' \wedge (x' \wedge y \vee y') \wedge (x \vee y) \wedge (x \vee y') =$}
\mbox{$x' \wedge y' \wedge (x \vee y) = 0$}.
\end{proof}

\begin{theorem} \label{the:ortho}
Properties GCC, CA, LCC, Z, N, LM, NP, RNL, LNL and NN are valid
in Orthomodular Quantic Logic and therefore in Hilbert Space Quantic Logic.
\end{theorem}
\begin{proof}
Let us show now that GCC holds.
Assume \mbox{$x * y \leq x$}. We shall show that  \mbox{$x * y = y * x$}.
First, note that, by Lemma~\ref{le:orth_thesis2}, 
\mbox{$x * y \leq$} \mbox{$x * (x \wedge y) =$}
\mbox{$x \wedge y \leq$} \mbox{$(x \vee y') \wedge y$}.
Therefore \mbox{$x * y \leq$} \mbox{$y * x$}.
By orthonormality, now, it is enough to prove that 
\mbox{$(x * y)' \wedge (y * x) = 0$}, i.e.,
\mbox{$(x * y)' \wedge (y \vee x') \wedge x = 0$}.
But \mbox{$( x * y)' \wedge x =$}
\mbox{$x \wedge (x' \wedge y \vee y') \leq$}
\mbox{$(x \vee y') \wedge (x' \wedge y \vee y' =$}
\mbox{$y'$} by orthonormality, since \mbox{$y' \leq x \vee y'$}.
Therefore \mbox{$( x * y)' \wedge x \leq$} \mbox{$x \wedge y'$}
and 
\[
(x * y)' \wedge (y \vee x') \wedge x \leq x \wedge y' \wedge (x' \vee y) =
(x' \vee y)' \wedge (x' \vee y) = 0.
\]
Let us show that CA holds.
Assume \mbox{$x * y =$} \mbox{$y * x$}.
We have \mbox{$x * y \leq$} \mbox{$x \wedge y \leq$} \mbox{$x * y$}.
Therefore \mbox{$x * y =$} \mbox{$x \wedge y$}.
We have \mbox{$z * (x * y) =$} \mbox{$z * (x \wedge y)$}.
By Lemma~\ref{le:orth_thesis3}, then, 
\mbox{$z * (x * y) =$} \mbox{$(z * x) * (x \wedge y) =$}
\mbox{$((z * x) * y) * (x \wedge y)$}.
But \mbox{$((z * x) * y \leq y$} and
\mbox{$((z * x) * y \leq$} \mbox{$x * y$} \mbox{$= y * x \leq x$} 
and therefore
\mbox{$((z * x) * y \leq$} \mbox{$x \wedge y$} and, as noticed above,
\mbox{$((z * x) * y) * (x \wedge y) =$} \mbox{$(z * x) * y$}.

The property LCC follows directly from Lemma~\ref{le:orth_thesis4}.
Properties Z, N, LM and NP are obvious.

Let us show that RNL holds.
Assume \mbox{$x * z \leq y$} and \mbox{$x * z' \leq y$}.
By orthomodularity: \mbox{$x \leq ((x \vee z') \wedge z) \vee z'$}
and therefore \mbox{$x \leq y \vee z'$}.
Also \mbox{$x \leq$} \mbox{$((x \vee z) \wedge z') \vee z$}
and therefore \mbox{$x \leq y \vee z$}.
Therefore \mbox{$x \leq (y \vee z) \wedge (y \vee z') \leq y$}.

Let us show that LNL holds.
By Lemma~\ref{le:orth_thesis5}, \mbox{$(z * x) \vee (z' * x) =$}
\mbox{$1 * x = x$}.
But, by assumption: \mbox{$(z * x) \vee (z' * x) \leq y$}.

Let us show that NN holds. By Lemma~\ref{le:orth_thesis6}
\mbox{$x \vee (x * z') =$} \mbox{$x \vee ((x \vee z) \wedge z') =$}
\mbox{$(x \vee z') \wedge (x \vee z)$}.
We see that \mbox{$x * z =$} \mbox{$(x \vee z') \wedge z \leq$}
\mbox{$(x \vee z') \wedge (x \vee z) =$} \mbox{$x \vee (x * z')$}.

Let us show that F4 holds. 
\mbox{$y * (x * y)' =$} \mbox{$(y \vee (x * y)) \wedge (x * y)' =$}
\mbox{$y \wedge (x * y)' =$} \mbox{$y \wedge ((x' \wedge y) \vee y'$}.
By Orthonormality this last expression is less or equal $x'$.
\end{proof}

\section{Ideals and a homomorphism theorem} \label{sec:homomorphism}
In this section, we generalize the notions of homomorphisms, kernels
and ideals that are fundamental in the study of Boolean algebras. We
prove a generalized homomorphism theorems: in non-commutative algebras
kernels and ideals coincide.  
\begin{definition} \label{def:morphism}
Let $S_{i}$, $i = 1,2$ be structures of the type considered in 
Section~\ref{sec:ncb-alg} of carriers $M_{1}$ and $M_{2}$ respectively. 
A function \mbox{$f : M_{1} \longrightarrow M_{2}$} is
a homomorphism from $S_{1}$ to $S_{2}$ iff, for any \mbox{$x, y \in M_{1}$}:
\begin{enumerate}
\item \mbox{$f(0) = 0$},
\item \mbox{$f(1) = 1$},
\item \mbox{$f(\neg x) = \neg f(x)$}, and
\item \mbox{$f(x * y) = f(x) * f(y)$}.
\end{enumerate}
\end{definition}
\begin{definition} \label{def:congruence}
If $S$ is a structure of carrier $M$, a binary relation $\sim$ on $M$ is
said to be a congruence relation iff:
\begin{enumerate}
\item $\sim$ is an equivalence relation,
\item if \mbox{$x \sim y$} then \mbox{$\neg x \sim \neg y$},
\item if \mbox{$x_{1} \sim x_{2}$} and \mbox{$y_{1} \sim y_{2}$} then
\mbox{$x_{1} * y_{1} \sim x_{2} * y_{2}$}.  
\end{enumerate}
\end{definition}
Any homomorphism $f$ defines a congruence relation \mbox{$\sim_{f}$}
by \mbox{$x \sim_{f} y$} iff \mbox{$f(x) = f(y)$}.
The kernel of $f$, $Ker(f)$ is the equivalence class of $0$.
We shall now study the relation between kernels and congruences.

The notion of an ideal is key. We need the following definition.
\begin{definition} \label{def:<=sim}
Assume $M$ is the carrier of a structure and 
\mbox{$I \subseteq M$}. We shall define two binary relations on $M$:
\begin{enumerate}
\item \mbox{$x \leq_{I} y$} iff \mbox{$x * \neg y \in I$}, and
\item \mbox{$x \sim_{I} y$} iff \mbox{$x \leq_{I} y$} and 
\mbox{$y \leq_{I} x$}.
\end{enumerate}
\end{definition}

\begin{definition} \label{def:ideal}
Assume \mbox{$S = \langle M, 0, 1, \neg, * \rangle$} is an NCNAB-algebra.
A set \mbox{$I \subseteq M$} is an {\em ideal} of S iff, for any
\mbox{$x, y, z \in M$}:
\begin{enumerate}
\item \label{zI} \mbox{$0 \in I$},
\item \label{*I} if \mbox{$x \in I$} then, for any \mbox{$y \in M$}
\mbox{$x * y \in I$} and \mbox{$y * x \in I$},
\item \label{vI} for any \mbox{$x, y, z \in M$}, if \mbox{$x * y \in I$}
and \mbox{$z * \neg y \in I$} then \mbox{$x * z \in I$}.
\item \label{*com} if \mbox{$(x * y) \leq{I} x$} then 
\mbox{$x * y \sim_{I} y * x$},
\item \label{*assoc} if \mbox{$x * y \sim_{I} y * x$}, 
then for any \mbox{$z \in M$}, \mbox{$(z * x) * y \sim_{I} z * (x * y)$},
\item \label{*lcom} if \mbox{$(z * x) * y \leq_{I} x$} and
\mbox{$(z * y) * x \leq_{I} y$}, then 
\mbox{$(z * x) * y \sim_{I} (z * y) * x$},
\end{enumerate}
\end{definition}
Condition~\ref{vI} corresponds to the Boolean condition: if $x$ and $y$ are
in $I$, then $x \vee y$ is in $I$: if $x$ and $y$ are in $I$, then
$(x \vee y) \wedge x$ and $(x \vee y) \wedge \neg x$ are in $I$ and therefore
$x \vee y$ is in $I$. 
Conditions~\ref{*com}, \ref{*assoc} and~\ref{*lcom} deal 
with the non-commutativity of $*$: they are trivially satisfied in a 
commutative structure.

\begin{lemma} \label{le:prelim_ideal}
Assume $I$ is an ideal.
\begin{enumerate}
\item if \mbox{$x \in I$} and \mbox{$y \leq x$} then \mbox{$y \in I$},
\item if \mbox{$x * y \in I$}, then \mbox{$y * x \in I$}.
\end{enumerate}
\end{lemma}
\begin{proof}
\begin{enumerate}
\item  By assumption and~\ref{*I}) we have \mbox{$y * x \in I$}.
But \mbox{$y * x = y$}.
\item
We have \mbox{$y * \neg y = 0 \in I$} and
\mbox{$x * y \in I$}. By~\ref{vI}) above 
\mbox{$y * x \in I$}.
\end{enumerate}
\end{proof}

\begin{lemma} \label{le:kernel}
Let $S_{1}$ be an NCNAB-algebra and $f$ is a homomorphism of domain $S{1}$, 
then, its kernel is an ideal.
\end{lemma}
\begin{proof}
\begin{enumerate}
\item By definition of a morphism \mbox{$f(0) = 0$}.
\item \mbox{$f(x) = 0$} implies \mbox{$f(x * y) =$}
\mbox{$f(x) * f(y) =$} \mbox{$0 * f(y) =$} $0$ and also
\mbox{$f(y * x) =$}
\mbox{$f(y) * f(x) =$} \mbox{$f(y) * 0 =$} $0$.
Note that \mbox{$x \leq_{I} y$} iff \mbox{$x * \neg y \in I$} iff
\mbox{$f(x) * \neg f(y) =$} $0$ iff \mbox{$f(x) \leq$} \mbox{$f(y)$}.
Also \mbox{$x \sim_{I} y$} iff \mbox{$f(x) =$} \mbox{$f(y)$}.
\item Assume \mbox{$f(x) * f(y) \leq f(x)$}.
By GCC, \mbox{$f(y) * f(x) =$}
\mbox{$f(x) * f(y) \leq$} \mbox{$f(y)$} and our conclusions hold.
\item Assume \mbox{$f(x) * f(y) =$} \mbox{$f(y) * f(x)$}.
By CA \mbox{$(f(z) * f(x)) * f(y) =$} \mbox{$f(z) * (f(x) * f(y))$}.
\item By LCC.
\item Assume \mbox{$f(x) * f(y) = 0$} and \mbox{$f(z) * \neg f(y) = 0$}.
We know that $f(x)$ and $f(y)$ commute and also that \mbox{$f(z) \leq f(y)$}.
Therefore \mbox{$f(x) * f(z) =$} \mbox{$(f(x) * f(y)) * f(z) =$} $0$.
\end{enumerate}
\end{proof}

\begin{lemma} \label{le:con_prelim}
If $I$ is an ideal of an NCNAB-algebra of carrier $M$, 
we have:
\begin{enumerate}
\item \label{<=} if \mbox{$x \leq y$} then \mbox{$x \leq_{I} y$},
\item \label{trans} the relation $\leq_{I}$ is transitive, and a quasi-order,
\item \label{equiv} the relation $\sim_{I}$ is an equivalence relation,
\item \label{comp} if \mbox{$x \leq_{I} y$} then 
\mbox{$\neg y \leq_{I} \neg x$}, and
\item \label{left<=} if \mbox{$x \leq_{I} y$} then, for any $z$
\mbox{$x * z \leq_{I} y * z$},
\item \label{<=sim} \mbox{$x \leq_{I} y$} iff \mbox{$x * y \sim_{I} x$},
\item \label{neg_com} if \mbox{$x * y \sim_{I} y * x$} then 
\mbox{$\neg x * y \sim_{I} y * \neg x$}.
\end{enumerate}
\end{lemma}
\begin{proof}
\begin{enumerate}
\item
If \mbox{$x \leq y$}, then \mbox{$x * \neg y = 0 \in I$}.
\item
Assume \mbox{$x * \neg y \in I$} and \mbox{$y * \neg z \in I$}.
By Lemma~\ref{le:prelim_ideal}
we have \mbox{$\neg z * y \in I$} and by
Definition~\ref{def:ideal}, item~\ref{vI}) \mbox{$x * \neg z \in I$}.
Since the relation $\leq_{I}$ is clearly reflexive by item~\ref{<=}) above
it is a quasi-order.
\item The relation is reflexive and transitive, by the above. It is symmetric
by definition.
\item Assume \mbox{$x * \neg y \in I$}. Then, by Lemma~\ref{le:prelim_ideal}
\mbox{$\neg y * x \in I$} and \mbox{$\neg y * \neg \neg x \in I$}.
Therefore \mbox{$\neg y * \neg \neg x \in I$}
and \mbox{$\neg y \leq_{I} \neg x$}.
\item Assume \mbox{$x * \neg y \in I$}. Since $z$ and 
\mbox{$\neg (y * z)$} commute, 
we have
\mbox{$(x * z) * \neg (y * z) =$} \mbox{$x * (z * \neg (y * z))$}.
But \mbox{$z * \neg (y * z) \leq \neg y$}. 
Therefore \mbox{$(x * z) * \neg (y * z) =$} 
\mbox{$(x * \neg y) * (z * \neg (y * z)) \in I$}.
\item Assume \mbox{$x * y \sim_{I} x$}. We have 
\mbox{$x \leq_{I} x * y \leq y$}. By parts~\ref{<=}) and~\ref{trans}) above
we conclude that \mbox{$x \leq_{I} y$}.
Assume \mbox{$x \leq_{I} y$}. We have \mbox{$y * x \leq x \leq_{I} y$}.
By parts~\ref{<=}) and~\ref{trans}) above
we conclude that \mbox{$y * x \leq_{I} y$}. By Definition~\ref{def:ideal},
part~\ref{*com}) then \mbox{$x * y \sim_{I} y * x$} and 
\mbox{$x * y \leq_{I} x$}. But \mbox{$x \leq_{I} y$} and, 
by~\ref{left<=} above, \mbox{$x * x \leq_{I} y * x \leq_{I} x * y$}.
We conclude that \mbox{$x \leq_{I} x * y$}.
\item Assume \mbox{$x * y \sim_{I} y * x$}.
By Definition~\ref{def:ideal}, part~\ref{*lcom}), 
\mbox{$(\neg x * x) * y \sim_{I}$} \mbox{$(\neg x * y) * x$}.
We see that \mbox{$(\neg x * y) * x \sim_{I} 0$}, i.e.,
\mbox{$(\neg x * y) * x \in I$}, i.e., \mbox{$\neg x * y \leq_{I} \neg x$}.
We conclude by Definition~\ref{def:ideal}, part~\ref{*com}) that
\mbox{$\neg x * y \sim_{I} y * \neg x$}.
\end{enumerate}
\end{proof}

\begin{theorem} \label{the:congruence}
If $I$ is an ideal of an NCNAB-algebra of carrier $M$, 
then the binary relation $\sim_{I}$ is a congruence relation.
\end{theorem}
\begin{proof}
The relation $\sim_{I}$ is obviously symmetric.
By Lemma~\ref{le:con_prelim} it is easily seen to be reflexive and transitive.
It is therefore an equivalence relation.
By Lemma~\ref{le:con_prelim}, \ref{comp}) \mbox{$x \sim_{I} y$}
implies \mbox{$\neg x \sim_{I} \neg y$}.
If \mbox{$x \sim_{I} y$}, then \mbox{$x * z \sim_{I}$} \mbox{$y * z$}
by  Lemma~\ref{le:con_prelim}, \ref{left<=}).

Assume now that \mbox{$x \sim_{I} y$}. We want to prove that
\mbox{$z * x \sim_{I}$} \mbox{$z * y$}.
It is enough to prove \mbox{$z * x \leq_{I}$} \mbox{$z * y$}.
We have \mbox{$z * y \leq$} \mbox{$y \leq_{I}$} $x$ and therefore, 
by~\ref{<=}) and~\ref{trans}) above:
\mbox{$z * y \leq_{I} x$}.
By Definition~\ref{def:ideal}, item~\ref{*com}),
\mbox{$(z * y) * x \sim{I}$} \mbox{$x * (z * y)$}.
By Lemma~\ref{le:con_prelim}, part~\ref{neg_com})
\mbox{$\neg (z * y) * x \sim_{I}$} \mbox{$x * \neg (z * y)$}.
By Definition~\ref{def:ideal} 
\mbox{$(z * x) * \neg (z * y) \sim_{I}$} 
\mbox{$z * (x * \neg (z * y))$}.
But \mbox{$x \leq_{I} y$} and therefore, by part~\ref{left<=}) above,
\mbox{$x * \neg (z * y) \leq_{I}$} \mbox{$y * \neg (z * y)$}.
By F4 \mbox{$y * \neg (z * y) \leq \neg z$} and therefore 
\mbox{$z * (x * \neg (z * y)) = 0$}.
We see that \mbox{$(z * x) * \neg (z * y) \sim_{I} 0$}.
We conclude that \mbox{$(z * x) * \neg (z * y) \in I$}
and \mbox{$(z * x) \leq_{I}$} \mbox{$z * y$}.
\end{proof}

Note that the converse of Lemma~\ref{le:kernel} holds. 
Any ideal is the kernel of some homomorphism.
\begin{theorem} [The homomorphism theorem] \label{the:homomorph}
If $I$ is an ideal, then it is the kernel of some homomorphism $f$
that is onto.
\end{theorem}
\begin{proof}
By Theorem~\ref{the:congruence}, the relation $\sim_{I}$ 
is a congruence relation. The operations $\neg$ and $*$ may 
therefore be defined on the set of equivalence classes under $\sim_{I}$ in
the natural way and $f$ defined by \mbox{$f(x) = \bar{x}$} is a homomorphisms
($\bar{x}$ is the equivalence class of $x$ under $\sim_{I}$).
One easily sees that the kernel of $f$ is $I$.
\end{proof}

\section{Future Work} \label{sec:future}
Here is a list of open questions and lines of enquiry.
\begin{itemize}
\item Are the properties of Theorems~\ref{the:star}, \ref{the:negation} 
and~\ref{the:misc} independent?
\item Do they characterize Hilbert Space Quantum Logic?
\item Find other structures that define NCNAB-algebras.
\item Find representation theorems for NCNAB-algebras, generalizing
known such results for Boolean algebras.
\item Consider operations that can be defined using $\neg$ and $*$. 
For example, \mbox{$\neg((x * \neg y) * (\neg x * y))$} seems to provide
a commutative exclusive disjunction.
\item Consider introducing additional operations in the syntax.
For example an implication that would be material implication in 
Boolean algebras and Sasaki hook in Hilbert space satisfying
\mbox{$z \leq x \rightarrow y$} iff \mbox{$z * x \leq y$}, 
or a disjunction satisfying
\mbox{$z * (x \vee y) \leq w$} iff \mbox{$z * x \leq w$} and
\mbox{$z * y \leq w$}.
\item What is the {\em right} definition of morphisms between P-families?
\item Do those morphisms preserve the lattice structure of the underlying
Hilbert spaces? 
\end{itemize}

\section{Acknowledgements} \label{sec:ack}
During the preparation of this work, the programs
for checking conjectures on orthomodular lattices available on 
Norman D. Megill's web site at
\newline
ftp://users.shore.net/members/n/d/ndm/quantum-logic 
have been an invaluable help. 
This work is a direct descendant of previous joint work 
with Kurt Engesser and Dov Gabbay. It also benefited from conversations
and arguments with them and with Alexei Grinbaum. 
I want to take this opportunity to thank them
warmly. Semyon Alesker, Joseph Bernstein and Vitali Milman were kind enough
to help me sort out my ideas on Hilbert spaces.

\bibliographystyle{plain}

\begin{thebibliography}{1}

\bibitem{BirkvonNeu:36}
Garret Birkhoff and John von Neumann.
\newblock The logic of quantum mechanics.
\newblock {\em Annals of Mathematics}, 37:823--843, 1936.

\bibitem{EngGabbay:Quantum}
K.~Engesser and D.M. Gabbay.
\newblock Quantum logic, {Hilbert} space, revision theory.
\newblock {\em Artificial Intelligence}, 136(1):61--100, March 2002.

\bibitem{Finch_lattice:69}
P.~D. Finch.
\newblock On the lattice structure of quantum logic.
\newblock {\em Bulletin of the Australian Mathematical Society}, 1:333--340,
  1969.

\bibitem{Greechie_strong:81}
R.~J. Greechie.
\newblock A non-standard quantum logic with a strong set of states.
\newblock In E.~G. Beltrametti and B.~C. van Fraassen, editors, {\em Current
  Issues in Quantum Logic}, volume~8 of {\em Ettore Majorana International
  Science Series}, pages 375--380. Plenum, New York, 1981.

\bibitem{KLMAI:89}
Sarit Kraus, Daniel Lehmann, and Menachem Magidor.
\newblock Nonmonotonic reasoning, preferential models and cumulative logics.
\newblock {\em Artificial Intelligence}, 44(1--2):167--207, July 1990.
\newblock CoRR: cs.AI/0202021.

\bibitem{LEG:Malg}
Daniel Lehmann, Kurt Engesser, and Dov~M. Gabbay.
\newblock Algebras of measurements: the logical structure of quantum mechanics.
\newblock {\em International Journal of Theoretical Physics}, 45(4):698--723,
  April 2006.
\newblock DOI 10.1007/s10773-006-9062-y.

\bibitem{McCune:Robbins}
William McCune.
\newblock Solution to the {R}obbins problem.
\newblock {\em Journal of Automated Reasoning}, 19(3):263--276, December 1997.
\newblock DOI 10.1023/A:1005843212881.

\bibitem{Roman_Rumbos:91}
L.~Rom\'{a}n and B.~Rumbos.
\newblock Quantum logic revisited.
\newblock {\em Foundations of Physics}, 21(6):727--734, June 1991.

\bibitem{vonNeumann:Quanten}
John von Neumann.
\newblock {\em Mathematische Grundlagen der Quanten-mechanik}.
\newblock Springer Verlag, Heidelberg, 1932.
\newblock American edition: Dover Publications, New York, 1943.

\end{thebibliography}

\end{document}